\documentclass[a4paper,10pt]{article}

\usepackage{float}
\usepackage[a4paper, left=0.5in, right=0.5in]{geometry}

\usepackage{latexsym}
\usepackage[utf8x]{inputenc}
\usepackage[english]{babel}
\usepackage{layout}
\usepackage{fancyhdr}
\usepackage{graphicx}
\usepackage{statex2}
\usepackage{caption}

\usepackage{subcaption}
\usepackage[]{graphicx}
\usepackage{float}

\usepackage{amsmath}
\usepackage{mathtools}
\usepackage{amsfonts}

\usepackage{booktabs}

\usepackage[round]{natbib}

\usepackage[font=footnotesize, labelfont={sf,bf}, margin=1cm]{caption}

\usepackage{amsmath}
\usepackage{mathtools}
\usepackage{amsfonts}

 \usepackage{amsthm}

\theoremstyle{plain}
\newtheorem{thm}{Theorem}[section]

\newtheorem{prep}[thm]{Preposition}

\theoremstyle{definition}
\newtheorem{defn}[thm]{Definition}

\theoremstyle{remark}

\theoremstyle{definition}
\newtheorem{exe}[thm]{Example}

\usepackage{epstopdf}

\newcommand{\stableD} {stable distribution}

\usepackage{dsfont}

\usepackage{authblk}
\usepackage[T1]{fontenc}

\usepackage{collcell}
\usepackage[table]{xcolor}

\usepackage{collcell}
\usepackage[table]{xcolor}
\usepackage{xstring}

\newcommand{\ApplyFontColor}[1]{
        \IfDecimal{#1}{
        \ifdim #1 pt > .99999 pt
             \color{red}\textbf{#1}
        \else
             \color{black}{#1}
        \fi  }
        {  \color{black}{#1} }
}

\newcolumntype{R}{>{\collectcell\ApplyFontColor}r<{\endcollectcell}}
\newcolumntype{C}{>{\collectcell\ApplyFontColor}c<{\endcollectcell}}

\title{ \textbf{Value-at-Risk Diversification of $\alpha$-stable Risks: \\ The Tail-Dependence Puzzle} }
\author[*]{Umberto Cherubini}
\author[*]{Paolo Neri}

\affil[*]{School of Economics, Management and Statistics \protect\\ Department of Statistics, University of Bologna,
 \protect\\ Bologna, Italy}

\begin{document}

\maketitle

\begin{abstract}
   We consider the problem of risk diversification of $\alpha$-stable heavy tailed risks. We study the behaviour of the aggregated \emph{Value-at-Risk}, with particular reference to the impact of different tail dependence structures on the limits to diversification. We confirm the large evidence of sub-additivity violations, particularly for risks with low tail index values and positive dependence. So, reinsurance strategies are not allowed to exploit diversification gains, or only a very limited amount of them. Concerning the impact of tail dependence, we find the peculiar results that for high tail dependence levels the limits to diversification are uniformly lower for all the levels of dependence, and for all levels of $\alpha<2$. The result is confirmed as we move towards extreme points in the tail: in this case, we show that at some point in the tail the aggregated VaR becomes additive above some level of dependence, but this critical dependence level is lower for copulas with lower tail dependence.
\end{abstract}
\section{Introduction}

Starting with the seminal paper by Artzner et al. (1999), a long debate has developed concerning risk measures and diversification, with particular focus on the fact that \emph{Value-at-Risk} (VaR) may fail to represent the diversification of risks. Following this long debate, the trend of regulation has drifted in favour of alternative risk measures ensuring sub-additivity. So, in market risk regulation a measure resembling \emph{Expected Shortfall} (ES) is bound to replace VaR in the new FRTB (\emph{Fundamental Review of the Trading Book}). The trend has been much less clear-cut in the insurance regulation, where the new Solvency II framework still relies on the VaR measure, while the corresponding regulation holding in Switzerland uses \emph{ES}.

If one considers the nature of the \emph{ES} measure and the tail behaviour of some actuarial risks, it is quite straightforward to explain the different trend in insurance. In fact, since \emph{ES} is the expected loss in the tail, this measure can only exist if the first moment of the loss distribution is well defined in the first place. Unfortunately, existence of first moment is very often challenged in the world of actuarial risks. This is particularly true for catastrophic risks, for which the fat tail phenomenon is so extreme to destroy the integrability requirement for existence of the first moment. For these kinds of risk, a plausible shape that has been suggested for the distribution of losses is the $\alpha$-stable family. In particular, \citet{ibragimov2008insurance} [IJW] show that seismic risks and other risk events typical of the catastrophe insurance market can be described by $\alpha$-stable distributions, some with $\alpha$ parameter lower than $1$, so that the first moment is not defined. The same result is documented by \citet{chavez2006quantitative} for a number of operational risks.

Concerning diversification, the IJW  paper also shows that diversification dramatically fails, even in the case of independent risks, that is where diversification should be highest. Beyond the discussion on the property of the risk measures, this finding points to a real relevant problem of the catastrophe insurance market, that is the fact that putting catastrophic risks together dramatically increases the ruin probability of the reinsurance policy of such risks, even if such risks are independent. This adds an important issue in the debate about the possibility of a private reinsurance market for catastrophic risks versus insurance schemes based on public funds. This also raises the question of how the results extend to the case in which catastrophic risks are dependent. This extension was first provided by \citet{ibragimov2016heavy} using the technique of copula functions.

The contribution of this paper to this stream of research in twofold. First, we want to explore the shape of the relationship between a dependence measure of the risks and a sub-additivity measure of the VaR. Of course, since it is well known that perfectly dependent risks are additive, the IJW result implies that this relationship must be non monotone for heavy tailed $\alpha$-stable risks. Second, we want to explore whether this diversification failure is made more severe by the tail-dependence feature of the risks.

The plan of the paper is as follows. In section 2 we define the convolution $\alpha$-stable risks under general dependence structures represented by copulas. In section 3 we present our simulation analysis addressing the two issues described above. Section 4 concludes.

\section{Preliminaries: Convolution of $\alpha$-stable Risks}

In this section, we describe the model that will be used in our simulations. First, we assume a pair of $\alpha$-stable distributed risks. Second, we assume a general dependence structure represented by a copula function with different degrees of tail-dependence. Third, we define the convolution of the two risks. For the ease of the reader, we review here the main concepts needed to follow our simulation exercise.

\subsection{Stable Distributions}

 Stable distributions is a wide class of probability distributions often used to represent heavy tailed models, and heavy tails may be so extreme that even first and second moments may fail to be well-defined. In insurance applications, this class of distributions is sometimes applied to catastrophic risks. Such type of risks includes earthquakes, floods, wind damage, etc... In these cases stable distributions are used to create loss distribution functions to model rare events with huge losses. Here we recall definitions and basic useful properties of univariate stable distributions, referring the reader to \citet{samorodnitsky1997stable} and \citet{nolan2012stable} for a detailed treatment.

 \begin{defn}[Stable]
    A random variable $X$ is stable if for $X_1, X_2,..,X_n$ independent copies of $X$ and $ S= \sum\limits_{i=1}^n X_i$, exist a, b \ $ \in \mathds{R}$ by which holds $ S= a X + b$.
    The random variable is strictly stable if holds with $b = 0$.
 \end{defn}

 \noindent The stable family is denoted by $\mathcal{S}(\alpha, \beta, \gamma, \delta)$. It is uniquely defined by the four parameters $\alpha, \beta, \gamma,$  and $\delta$ that determine the density function. These parameters can be interpreted as follows:
    \begin{itemize}
      \item[($\alpha$)] is the basic stability parameter, determining the weight in the tails: the lower the $\alpha$ value, the greater the frequency and size of extreme events. Its range is $(0 < \alpha \leq 2)$

      \item[($\beta$)]  It is the skewness distribution parameter and  $(-1 \leq \beta \leq 1)$, with a zero value indicating that the distribution is symmetric. A negative / positive $\beta$ implies that the distribution is skewed to the left or right respectively

      \item[($\gamma$)] The parameter $\gamma$ is positive and represents a measure of dispersion.

      \item[($\delta$)] The parameter $\delta$ determines the location of the distribution.
    \end{itemize}

  Closed forms for $\alpha$-stable distributions generally do not exist, except for specific cases of the parameters pair$(\alpha, \beta)$.

  \qquad Lévy: $\mathcal{S}(0.5, 1, \gamma, \delta) \qquad f(x)= \sqrt{\frac{\gamma}{2 \pi}} \frac{1}{ (x- \delta)^{3/2}} \exp \Bigl(- \frac {\gamma}{2(x - \delta)} \Bigr), \qquad \delta \leq x \leq \infty $ \\

  \qquad Cauchy: $\mathcal{S}(1, 0, \gamma, \delta) \qquad f(x)= \frac{1}{\sqrt{\pi}\sigma} \frac{\gamma}{\gamma^2 + (x- \delta)^2}, \qquad -\infty \leq x \leq \infty $ \\

  \qquad Gaussian: $\mathcal{S}(2, 0, \sigma, \mu) \qquad f(x)= \frac{1}{\sqrt{2\pi}\sigma} \exp \Bigl(- \frac{(x - \mu)^2}{2\sigma^2} \Bigr), \qquad -\infty \leq x \leq \infty $ \\

 The most direct way to describe all possible stable distributions is through the characteristic function.
\\

\begin{defn}[Stable random variable] 
        A r.v. X is stable if and only if it is expressed by the following characteristic function $\varphi_X(t)$
        \begin{equation}\label{eq:stable}
            \mathbb{E} \Bigl[ \varphi_X(t) \Bigr] = \mathbb{E} \Bigl[ e^{\imath t X} \Bigr] =
            \left\{ \begin{aligned}
                 & \exp\{\imath \delta t - \gamma^{\alpha } |t|^{\alpha } (1 - \imath \beta sign(t) \tan(\pi \alpha/2)) \}, \: \alpha \neq 1  \\
                 & \exp\{\imath \delta t - \gamma |t| (1 + \imath \beta sign(t) \tfrac{2}{\pi} \ln|t| ) \}, \: \alpha = 1
            \end{aligned}
            \right.
        \end{equation}
\end{defn}
\begin{prep}
    The Family of \stableD\ $\mathcal{S}(\alpha, \beta, \gamma, \delta)$ is described by the following properties \\
    \begin{itemize}
      \item  [a)] Let X  \ $ \in \: \mathcal{S}(\alpha, \beta, \gamma, \delta), \: \forall a \neq 0 ,b \in \mathds{R}$
    \begin{equation} \label{eq:par_a_1}
      aX + b \sim
      \begin{cases}
        \mathcal{S}(\alpha, sign(\alpha) \beta, |a|\gamma, a \delta + b), \qquad \alpha \neq 1 \\
        \mathcal{S}(\alpha, sign(\alpha) \beta, |a| \gamma, a \delta  + b - \frac{2}{\pi} \beta \gamma a \log |a|), \qquad \alpha = 1
      \end{cases}
    \end{equation}

       \item [b)]  Let  $X_1  \in \: \mathcal{S}(\alpha, \beta_1, \gamma_1, \delta_1)$ e $X_2 \in \: \mathcal{S}(\alpha, \beta_2, \gamma_2, \delta_2)$ independent r.v's. The the sum $X_1+X_2 \sim \mathcal{S}(\alpha, \beta, \gamma, \delta) $ , where
    \begin{equation*}
        \beta = \frac{\beta_1 \gamma_1^\alpha + \beta_2 \gamma_2^\alpha}{\gamma_1^\alpha + \gamma_2^\alpha}, \quad
        \gamma= (\gamma_1^\alpha + \gamma_2^\alpha)^{1 / \alpha}, \quad
        \delta=  \delta_1 + \delta_2
    \end{equation*}

      \item [c)] For all values in  $0 < \alpha < 2$, it follows that
    \[ X \sim \mathcal{S}(\alpha, \beta, \gamma, 0) \Leftrightarrow  -X \sim \mathcal{S}(\alpha, -\beta, \gamma, 0) \]

      \item [d)] An important result involves p-order moments of a Stable r.v. Let X  $ \sim \mathcal{S}(\alpha, \beta, \gamma, 0), 0 < \alpha < 2$. It holds
    \begin{align*}
        \mathbb{E}|X|^p &< \infty , \qquad  \forall \; 0 < p < \alpha \\
        \mathbb{E}|X|^p &= \infty , \qquad  \forall \; p \geq \alpha
    \end{align*}

    \noindent It follows that the stable family does not admit finite second moment for $\alpha < 2$ and so variance does not exist, and if $\alpha \leq 1$ the first moment does not exist either.
    \end{itemize}

\end{prep}

\subsection{Copula functions and Dependence}
We recall the basic theory and useful results about copula functions and the non parametric dependence structure of random variables. A complete reference on copula functions is \citet{nelsen2007introduction}, while \citet{frees1998understanding} describes copula simulation methods. A reference on copula methods applied in finance is \citet{cherubini2004copula}.

\begin{defn}[Copula]
A n-dimensional Copula $\mathcal{C}$ is a function $ \mathcal{C} : U_1 \times U_2 \cdot \cdot \cdot \times U_n \rightarrow \mathds{I} ,  U_i  \in \mathds{I}$  the unit space,  $\mathds{I} \ \in (0,1)$,  with the following properties

\begin{enumerate}
  \item $ \forall i=1...n, \ \mathcal{C}(1,1,...,U_i,...,1,1)= U_i $
  \item $ \forall i=1...n, \ \mathcal{C}(U_1,U_2,...,U_i=0,...,U_{n-1},U_n)= 0 $
  \item $ \forall \ (u_1,...,u_n), (v_1,...,v_n) \ \in [0,1]^n  $ where $ u_i \leq v_i$, holds \\
        \begin{center}
            $ \sum\limits_{i_1=1}^2  \cdot \cdot \cdot \sum\limits_{i_n=1}^2 (-1)^{i_1,...,i_n} \mathcal{C}(s_{1,i_1},..., s_{n,i_n})\geq 0 $ \hspace{2em} where $s_{j,1}= u_j$ and $s_{j,2}= v_j$
        \end{center}
\end{enumerate}
\end{defn}

Copula functions allow to separate the marginal behaviour of individual random variables from their dependence structure. Due to the integral probability transform, copula functions single out the dependence structure by reducing the representation of dependence to a function taking uniform random variables as arguments.

\begin{thm}[Sklar]
    Let $ F_{i} \ \forall i=1...n $. be a set of continuous univariate distribution functions of the random variables $(X_1,X_2,\ldots,X_n)$\\ Then, $H(X_1,X_2,\ldots,X_n)$ is a joint distribution function if and only if $\forall (x_1, x_2,..., x_n) \in \mathds{R}^n$
    \begin{equation}
              H(x_1, x_2,..., x_n)= \mathcal{C}(F_{1},F_{2},..., F_{n}) \label{eq:Sklar}
    \end{equation}
    where $\mathcal{C}$ is a copula function.
\end{thm}

It is well known that copula functions are naturally linked to non parametric association measures such as Spearman's $\rho$ and {Kendall's} $\tau$.

\begin{prep}
Suppose (X,Y) have continuous marginal distributions and unique copula $\mathcal{C}$. Then the {Kendall's} $\tau$  are given by
    \begin{equation*}
         \rho_\tau(X,Y) = 4 \int_{0}^1\int_{0}^1{ \mathcal{C}(u,v) \,d\mathcal{C}(u,v) - 1 }
    \end{equation*}
\end{prep}

\noindent and rank correlation measure {Spearman's} $\rho$ is defined as
\begin{equation*}
       \rho_S(X,Y)= 12 \int_{0}^1\int_{0}^1{ \mathcal{C}(u,v) \,d\mathcal{C}(u,v)- 3 }
\end{equation*}

Copula functions are also useful to represent the dependence structure of extreme events, that is measured by the so called \emph{tail dependence} coefficients. These are linked to the concept of conditional distribution, and how this can be written in terms of copula functions. Formally, we have the following:

\begin{defn} \label{def:lambda}
Let X,Y be two  r.v's with corresponding continuous d.f.'s F, G. We define the upper tail dependence parameter $\lambda_U$  as the limit,
\begin{equation}
    \lambda_U= \lim_{t \to 1^-} \mathds{P} \Big[ Y > G^{(-1)}(t) \Big| X > F^{(-1)}(t) \Big]\label{eq:lambu}
\end{equation}
\noindent Accordingly, the lower tail dependence coefficient $\lambda_L$ is defined as
\begin{equation}
    \lambda_L= \lim_{t \to 0^+} \mathds{P} \Big[ Y \le G^{(-1)}(t) \Big| X \le F^{(-1)}(t) \Big]\label{eq:lambl}
\end{equation}
\end{defn}
\noindent Next theorem shows how tail dependence can be expressed in terms of copulas.
\begin{thm}\label{eq:teocoda}
     Given X, Y, F, G, $\lambda_U$, $\lambda_L$ defined as above in (\ref{def:lambda}), let $\mathcal{C}$ be the bivariate copula of X,Y. If the limits (\ref{eq:lambu}) and (\ref{eq:lambl}) exist, then tail dependence coefficients can be expressed as follows
    \begin{align*}
      \lambda_U&= 2 - \lim_{t \to 1^-}\frac{1-C(t,t)}{1-t} \\
      \lambda_L&= \lim_{t \to 0^+}\frac{C(t,t)}{t}
    \end{align*}
\end{thm}
\noindent Here below we show some tail dependence formulas for the most commonly used copulas: elliptical (gaussian and Student-t) and Archimedean (Clayton, Frank and Gumbel). We note that Student-t, Gumbel and Clayton copula show tail dependence.
\begin{exe}[Copulas and corresponding  tail dependencies]
\mbox{}\\*\indent
    \begin{itemize}
      \item  Gaussian copula   \scalebox{0.95}{
            $ C_\rho(u_1, u_2, ...,u_n)= \phi^n( \phi^{-1}(u_1), \phi^{-1}(u_2), ..., \phi^{-1}(u_n), \mathbf{\rho})  \qquad \lambda_U= \lambda_L= 0$
      }

      \item Student-t copula \scalebox{0.95}{
         $ C_\rho^\nu(u_1, u_2, ...,u_n)= t^n( t^{-1}(u_1), t^{-1}(u_2),..., t^{-1}(u_n),\mathbf{\rho},\nu) \qquad
          \lambda_U = \lambda_L = 2 t_{\nu+1} \left( \sqrt{\nu +1} \frac{\sqrt{1 - \rho}}{\sqrt{1 + \rho}} \right) $
      }

      \item Clayton copula \scalebox{0.95}{
            $ C_\theta(u,v)= [u^{-\theta} + v^{-\theta} -1]^{-\frac{1}{\theta}}  \qquad
            \lambda_L=\displaystyle{\lim_{t \to \infty} } \frac{\varphi^{-1}(2t)}{\varphi^{-1}(t)}   =   \displaystyle{\lim_{t \to \infty} } \frac{ (1+2t \theta)^{ - \frac{1}{\theta} } }{(1+t \theta)^{- \frac{1}{\theta}}} \: = \: 2^{-\frac{1}{\theta}}$
      }

      \item Frank copula \scalebox{0.95}{
            $ C_\theta(u,v)=  -\frac{1}{\theta} \ln \left( 1 +
            \frac{(\mathrm{e}^{-\theta u} -1)(\mathrm{e}^{-\theta v} -1)} {\mathrm{e}^{-\theta} -1} \right)\qquad $}
            \text{like the Gaussian} \scalebox{0.95}{$ \qquad  \lambda_U= \lambda_L= 0$}

      \item Gumbel copula \scalebox{0.95}{
            $ C_\theta(u,v)=  \mathrm {e}^{-[(- \ln u)^{-\theta} + (- \ln v)^{-\theta}]^{\frac{1}{\theta}}} \qquad
             \lambda_U = \displaystyle{\lim_{t \to 0^+ }} \frac{1 - \varphi^{-1}(2t)}{1 - \varphi^{-1}(t)} = \displaystyle{\lim_{t \to 0^+ } } \frac{ 1- \mathrm{e}^{-2t^{\frac{1}{\theta}}} }
            {1- \mathrm{e}^{-t^{\frac{1}{\theta}}}}  = 2 - 2^{\frac{1}{\theta}} $
      }
    \end{itemize}
\end{exe}

\noindent

\subsection{Convolution of dependent variables}

We now formally introduce the representation of the VaR aggregation problem for dependent variables. Under the assumption of independence, the problem of computing the VaR of a sum of risks would merely leads to the standard concept of convolution of random variables. Then, the natural extension of this concept to the case of dependent variables is the definition of \emph{$C$-convolution} proposed by \citet{cherubini2011dynamic}. The term $C$ in the definition reminds of the copula function representing a general relationship of a pair of variables $X$ and $Y$. The finding was obtained as a by-product of a more general result for the characterization of the dependence structure between the variables $X$ and $X+Y$ from the dependence between $X$ and $Y$. We report here the main proposition.

\begin{prep}
Let X,Y be two real-valued random variables on the same probability space $(\Omega, \mathcal{F}, \mathds{P})$ with corresponding copula $C_{X,Y}$ and continuous marginals $F_X, F_Y$. Then,
    \begingroup
        \addtolength{\jot}{0.8em}
        \begin{align}
          C_{X,X+Y}(u,v) &=  \displaystyle\int_{0}^{u} D_1 C_{X,Y}\Big(\omega, F_Y(F_{X+Y}^{-1}(v) - F_{X}^{-1}(\omega))\Big) d \omega \label{eq:t1}  \\
          F_{X+Y}(t)   &=  \displaystyle\int_{0}^{1} D_1 C_{X,Y}\Big(\omega, F_Y( t - F_{X}^{-1}(\omega))\Big) d \omega \label{eq:t2}
        \end{align}
    \endgroup
where $C_{i,j}(u,v)$ denotes the copula functions between the variables reported in the underscore,  $D_1C(u,v)$ represents the derivative with respect to $u$, $F_i$ denotes the distribution function of the variable reported in the underscore($X$, $Y$ and $X+Y$).
\end{prep}

Notice that in this result we have implicity defined te concept of  \textit{C-convolution}
\begin{defn}
    Lef $F$, $H$ be two continuous c.d.f's and  $C$ a copula function. The \textit{C-convolution} of $H$ and $F$ is defined as the c.d.f.
    \begin{equation*}
      H \overset{C}{\ast}  F(t) = \displaystyle\int_{0}^{1} D_1 C \Big(\omega, F(t - {H}^{-1}(\omega))\Big) \ d \omega
    \end{equation*}
\end{defn}

\noindent This result provides a formal definition and an alternative computation approach to the convolution of dependent variables, with respect to the standard Monte Carlo. In some cases, moreover, an important property of the \emph{C-convolution} operator can be used to make the computation easier. In fact, it can be easily proved that the \emph{C-convolution} operator is closed with respect to mixture of copula functions. In other words, it can be shown that if for some bivariate copula functions $A$ and $B$ we have
\begin{equation*}
  C(u,v)= \lambda A(u,v) + (1- \lambda) B(u,v) , \; \forall \ \lambda \ \in \ [0,1]
\end{equation*}
then, for all c.d.f's $H$, $F$ it holds
\begin{equation*}
  H \overset{C}{\ast}  F(t) = H \overset{\lambda A + (1- \lambda) B }{\ast} F = \lambda H \overset{A}{\ast} F + (1- \lambda) H \overset{B}{\ast} F
\end{equation*}

\section{Application}

In this section, we consider the problem of risk diversification in a VaR framework with heavy-tailed $\alpha$-stable risks and arbitrary dependence captured by a copula function. As motivated in the introduction, VaR may be the only available measure of risk in $\alpha$-stable cases when not even the first moment is defined, that is when $\alpha \leq 1$. We also remind that we are interested in exploring two issues:
\begin{itemize}
  \item the shape of the non-monotonic relationship between dependence and the degree of super-additivity
  \item the role played by tail-dependence in the degree of super-additivity.
\end{itemize}

We consider two identically distributed $\alpha$-stable r.v.s' $X, Y$ and their portfolio $S= (X+Y)$ with dependence structure represented by copulas with different tail-dependence. We use the Kendall-$\tau$ concordance measure to span a wide range of dependence levels. For each level, we compute a super-additivity index,  as in \citet{ibragimov2016heavy}

\begin{equation*}
  SR = \frac{VaR_q(X + Y)}{VaR_q(X) + VaR_q(Y)}
\end{equation*}

Super-additivity is detected if SR $ \geq 1 $, so that $VaR_q(X + Y)$, the risk corresponding to the portfolio, is greater than the sum of separated risks $VaR_q(X) + VaR_q(Y)$. When instead $ SR < 1 $, diversification is at work and the aggregation of risks reduces the probability of losing the capital allocated to support them.

\subsection{Simulation and general results}

The analysis was performed using the two methods available, that is by numerical integration of the $C-convolution$ formula and by standard Monte Carlo. Unfortunately, in some cases it was particularly difficult to apply numerical integration, so that for completeness of the report, here we only show the results used with Monte Carlo. So, we generated a portfolio of two assets $X, Y$ drawn from heavy-tailed marginal distribution with several tail indexes $\alpha$ ranging between $0 < \alpha \leq2$ and linked by a bivariate copula determining their dependence structure. Each simulation is based on $1,000,000$ draws.

The tables 1 to 5 report the results for the main copula functions typically used, that is the elliptical (gaussian and Student-t) and the Archimedean ones (Clayton, Gumbel and Frank). As it is well known, only the Gaussian and Frank copulas do not have tail dependence, the Gumbel has only upper tail dependence, the Clayton has only lower tail dependence ans Student-t has both upper and lower tail dependence.

All these results for the different copulas seem to design a consistent picture. Some diversification benefits typically appear in the upper and right parts of the tables. Subadditivity holds in cases in which dependence is negative and the value of the tail index $\alpha$ is higher than $1$ (this is also true for the Gumbel case in which negative dependence is not allowed). So, the failure of diversification is a general result for cases in which the first moment does not exist. The results are instead mixed in the region in which the first moment is defined, where diversification may still fail for higher level of dependence. Moreover, even in the positive dependence region in which diversification benefits materialize, their amount is of a limited order.

\begin{table}[H]
\centering
\caption{MonteCarlo simulation: Super-additivity Ratio, gaussian copula}
\label{my-Gaussian}
\makebox[\textwidth]{
\scalebox{0.75}{
\begin{tabular}{cc *{11}{C}}

\toprule
q=5\% & $\alpha$ & \multicolumn{1}{c}{\textbf{0.2}} & \multicolumn{1}{c}{\textbf{0.3}} & \multicolumn{1}{c}{\textbf{0.4}} & \multicolumn{1}{c}{\textbf{0.6}} & \multicolumn{1}{c}{\textbf{0.9}} & \multicolumn{1}{c}{\textbf{1}} & \multicolumn{1}{c}{\textbf{1.1}} & \multicolumn{1}{c}{\textbf{1.3}} & \multicolumn{1}{c}{\textbf{1.6}} & \multicolumn{1}{c}{\textbf{1.8}} & \multicolumn{1}{c}{\textbf{2}} \\
\midrule

\multicolumn{1}{c}{Param. $\theta$ } & \multicolumn{1}{c}{ Kendall-$\tau$} & \multicolumn{2}{r}{\textbf{Gaussian copula}} &  &   &  &   &   &   &   &   &   \\

\midrule

-1 & -1 & 5.0e-06 & 3.5e-06 & 2.7e-06 & 1.9e-06 & 1.3e-06 & 1.2e-06 & 1.1e-06 & 9.8e-07 & 8.8e-07 & 8.4e-07 & 7.9e-07 \\
-0.99 & -0.9 & 0.5443 & 0.3593 & 0.2754 & 0.1929 & 0.1368 & 0.1241 & 0.1142 & 0.0980 & 0.0878 & 0.0835 & 0.0785 \\
-0.89 & -0.7 & 2.4852 & 1.2751 & 0.8931 & 0.5832 & 0.4034 & 0.3676 & 0.3378 & 0.2920 & 0.2608 & 0.2483 & 0.2334 \\
-0.59 & -0.4 & 8.7629 & 3.2337 & 1.9524 & 1.1392 & 0.7621 & 0.6946 & 0.6416 & 0.5661 & 0.5063 & 0.4794 & 0.4543 \\
-0.45 & -0.3 & 11.3880 & 3.9143 & 2.2594 & 1.2959 & 0.8598 & 0.7878 & 0.7324 & 0.6497 & 0.5813 & 0.5505 & 0.5225 \\
-0.16 & -0.1 & 15.3299 & 4.8540 & 2.7295 & 1.5192 & 1.0226 & 0.9452 & 0.8810 & 0.7938 & 0.7136 & 0.6786 & 0.6493 \\
-0.02 & -0.01 & 16.0090 & 5.0211 & 2.8224 & 1.5814 & 1.0740 & 0.9935 & 0.9344 & 0.8469 & 0.7656 & 0.7291 & 0.7015 \\
0 & 0 & 15.8676 & 5.0450 & 2.8237 & 1.5858 & 1.0792 & 1.0009 & 0.9386 & 0.8522 & 0.7698 & 0.7350 & 0.7072 \\
0.02 & 0.01 & 16.1089 & 5.0292 & 2.8408 & 1.5927 & 1.0840 & 1.0043 & 0.9430 & 0.8570 & 0.7765 & 0.7406 & 0.7127 \\
0.16 & 0.1 & 15.6895 & 4.9868 & 2.8299 & 1.6148 & 1.1193 & 1.0417 & 0.9832 & 0.9015 & 0.8215 & 0.7865 & 0.7604 \\
0.31 & 0.2 & 14.0857 & 4.7032 & 2.7219 & 1.5978 & 1.1398 & 1.0676 & 1.0132 & 0.9409 & 0.8661 & 0.8323 & 0.8088 \\
0.45 & 0.3 & 11.9989 & 4.2562 & 2.5444 & 1.5526 & 1.1439 & 1.0826 & 1.0343 & 0.9672 & 0.9032 & 0.8724 & 0.8528 \\
0.59 & 0.4 & 9.5069 & 3.6756 & 2.3091 & 1.4737 & 1.1361 & 1.0804 & 1.0418 & 0.9869 & 0.9332 & 0.9070 & 0.8909 \\
0.89 & 0.7 & 3.2120 & 1.8769 & 1.4633 & 1.1810 & 1.0556 & 1.0358 & 1.0230 & 1.0053 & 0.9870 & 0.9773 & 0.9725 \\
0.99 & 0.9 & 1.2798 & 1.1202 & 1.0652 & 1.0234 & 1.0072 & 1.0047 & 1.0038 & 1.0011 & 0.9990 & 0.9973 & 0.9969 \\
1 & 1 & 1 & 1.0000 & 1.0000 & 1 & 1.0000 & 1.0000 & 1.0000 & 1.0000 & 1.0000 & 1.0000 & 1 \\

\bottomrule
\end{tabular}}}
\end{table}

\begin{table}[H]
\centering
\caption{MonteCarlo simulation: Super-additivity Ratio, Student-t$(\nu=5)$ copula}
\label{my-Student}
\makebox[\textwidth]{
\scalebox{0.75}{
\begin{tabular}{cc *{11}{C}}

\toprule
q=5\% & $\alpha$ & \multicolumn{1}{c}{\textbf{0.2}} & \multicolumn{1}{c}{\textbf{0.3}} & \multicolumn{1}{c}{\textbf{0.4}} & \multicolumn{1}{c}{\textbf{0.6}} & \multicolumn{1}{c}{\textbf{0.9}} & \multicolumn{1}{c}{\textbf{1}} & \multicolumn{1}{c}{\textbf{1.1}} & \multicolumn{1}{c}{\textbf{1.3}} & \multicolumn{1}{c}{\textbf{1.6}} & \multicolumn{1}{c}{\textbf{1.8}} & \multicolumn{1}{c}{\textbf{2}} \\
\midrule

\multicolumn{1}{c}{Param. $\theta$ } & \multicolumn{1}{c}{ Kendall-$\tau$} & \multicolumn{2}{r}{\textbf{Student-t copula {$(\nu=5)$} }} &  &   &  &   &   &   &   &   &   \\

\midrule

-1 & -1 & 4.2e-06 & 2.9e-06 & 2.3e-06 & 1.7e-06 & 1.2e-06 & 1.1e-06 & 1e-06 & 9.3e-07 & 8.6e-07 & 8.2e-07 & 7.7e-07 \\
-0.99 & -0.9 & 0.4496 & 0.3004 & 0.2311 & 0.1658 & 0.1198 & 0.1104 & 0.1027 & 0.0927 & 0.0858 & 0.0816 & 0.0769 \\
-0.89 & -0.7 & 1.8897 & 1.0421 & 0.7376 & 0.4997 & 0.3557 & 0.3281 & 0.3061 & 0.2763 & 0.2550 & 0.2429 & 0.2294 \\
-0.59 & -0.4 & 6.1498 & 2.5362 & 1.6017 & 0.9842 & 0.6844 & 0.6331 & 0.5929 & 0.5358 & 0.4920 & 0.4700 & 0.4484 \\
-0.45 & -0.3 & 7.9169 & 3.0426 & 1.8669 & 1.1213 & 0.7806 & 0.7217 & 0.6767 & 0.6141 & 0.5638 & 0.5401 & 0.5172 \\
-0.16 & -0.1 & 10.5219 & 3.7700 & 2.2448 & 1.3357 & 0.9367 & 0.8715 & 0.8222 & 0.7521 & 0.6935 & 0.6670 & 0.6446 \\
-0.02 & -0.01 & 10.9465 & 3.8993 & 2.3270 & 1.3924 & 0.9894 & 0.9217 & 0.8736 & 0.8050 & 0.7456 & 0.7187 & 0.6970 \\
0.00 & 0 & 11.0378 & 3.9264 & 2.3414 & 1.3982 & 0.9942 & 0.9271 & 0.8787 & 0.8102 & 0.7511 & 0.7241 & 0.7028 \\
0.02 & 0.01 & 11.0536 & 3.9473 & 2.3387 & 1.4050 & 0.9961 & 0.9327 & 0.8839 & 0.8157 & 0.7561 & 0.7296 & 0.7088 \\
0.16 & 0.1 & 10.6819 & 3.9086 & 2.3511 & 1.4276 & 1.0338 & 0.9721 & 0.9250 & 0.8599 & 0.8015 & 0.7760 & 0.7572 \\
0.31 & 0.2 & 9.9593 & 3.7179 & 2.2905 & 1.4310 & 1.0600 & 1.0041 & 0.9607 & 0.9005 & 0.8466 & 0.8225 & 0.8059 \\
0.45 & 0.3 & 8.5283 & 3.3891 & 2.1585 & 1.3972 & 1.0739 & 1.0241 & 0.9848 & 0.9332 & 0.8860 & 0.8641 & 0.8503 \\
0.59 & 0.4 & 6.8569 & 2.9793 & 1.9817 & 1.3481 & 1.0743 & 1.0328 & 1.0005 & 0.9585 & 0.9195 & 0.9002 & 0.8894 \\
0.89 & 0.7 & 2.6423 & 1.6706 & 1.3521 & 1.1307 & 1.0340 & 1.0180 & 1.0087 & 0.9957 & 0.9820 & 0.9752 & 0.9717 \\
0.99 & 0.9 & 1.2236 & 1.0999 & 1.0502 & 1.0177 & 1.0045 & 1.0028 & 1.0018 & 0.9997 & 0.9982 & 0.9972 & 0.9968 \\
1 & 1 & 1.0000 & 1 & 1.0000 & 1 & 1 & 1.0000 & 1 & 1.0000 & 1.0000 & 1.0000 & 1.0000 \\

\bottomrule
\end{tabular}}}
\end{table}
\begin{table}[H]
\centering
\caption{MonteCarlo simulation: Super-additivity Ratio, Frank copula}
\label{my-Frank}
\makebox[\textwidth]{
\scalebox{0.75}{
\begin{tabular}{cc *{11}{C}}

\toprule
q=5\% & $\alpha$ & \multicolumn{1}{c}{\textbf{0.2}} & \multicolumn{1}{c}{\textbf{0.3}} & \multicolumn{1}{c}{\textbf{0.4}} & \multicolumn{1}{c}{\textbf{0.6}} & \multicolumn{1}{c}{\textbf{0.9}} & \multicolumn{1}{c}{\textbf{1}} & \multicolumn{1}{c}{\textbf{1.1}} & \multicolumn{1}{c}{\textbf{1.3}} & \multicolumn{1}{c}{\textbf{1.6}} & \multicolumn{1}{c}{\textbf{1.8}} & \multicolumn{1}{c}{\textbf{2}} \\
\midrule

\multicolumn{1}{c}{Param. $\theta$ } & \multicolumn{1}{c}{ Kendall-$\tau$} & \multicolumn{2}{r}{\textbf{Frank copula} } &  &   &  &   &   &   &   &   &   \\

\midrule

-709 & -1 & 0.0698 & 0.0483 & 0.0376 & 0.0261 & 0.0181 & 0.0163 & 0.0147 & 0.0120 & 0.0088 & 0.0074 & 0.0065 \\
-38.28 & -0.9 & 2.3807 & 1.1773 & 0.7939 & 0.4945 & 0.3189 & 0.2846 & 0.2555 & 0.2065 & 0.1507 & 0.1257 & 0.1088 \\
-11.41 & -0.7 & 8.4532 & 3.1117 & 1.8384 & 1.0425 & 0.6610 & 0.5911 & 0.5312 & 0.4402 & 0.3435 & 0.3029 & 0.2727 \\
-4.16 & -0.4 & 13.5038 & 4.4572 & 2.4888 & 1.3722 & 0.8904 & 0.8077 & 0.7394 & 0.6404 & 0.5492 & 0.5088 & 0.4756 \\
-2.92 & -0.3 & 14.7779 & 4.6550 & 2.6111 & 1.4407 & 0.9428 & 0.8596 & 0.7968 & 0.6990 & 0.6102 & 0.5701 & 0.5383 \\
-0.91& -0.1 & 15.8484 & 4.9234 & 2.7937 & 1.5473 & 1.0357 & 0.9593 & 0.8947 & 0.8046 & 0.7215 & 0.6845 & 0.6553 \\
-0.09 & -0.01 & 15.9420 & 5.0413 & 2.8390 & 1.5899 & 1.0759 & 0.9937 & 0.9357 & 0.8467 & 0.7656 & 0.7316 & 0.7014 \\
    0 & 0 & 16.0998 & 5.0095 & 2.8282 & 1.5900 & 1.0811 & 0.9993 & 0.9373 & 0.8514 & 0.7709 & 0.7347 & 0.7071 \\
0.09 & 0.01 & 15.9468 & 5.0584 & 2.8299 & 1.5919 & 1.0835 & 1.0070 & 0.9424 & 0.8582 & 0.7769 & 0.7392 & 0.7113 \\
0.91 & 0.1 & 15.8797 & 5.0614 & 2.8471 & 1.6093 & 1.1162 & 1.0421 & 0.9804 & 0.8948 & 0.8162 & 0.7812 & 0.7546 \\
1.86 & 0.2 & 15.7188 & 5.0186 & 2.8649 & 1.6352 & 1.1483 & 1.0732 & 1.0178 & 0.9373 & 0.8604 & 0.8240 & 0.7987 \\
2.92 & 0.3 & 15.0588 & 4.9029 & 2.8181 & 1.6495 & 1.1761 & 1.1040 & 1.0505 & 0.9729 & 0.8979 & 0.8626 & 0.8370 \\
4.16 & 0.4 & 14.2756 & 4.7217 & 2.7806 & 1.6555 & 1.1998 & 1.1306 & 1.0806 & 1.0088 & 0.9337 & 0.8970 & 0.8730 \\
11.41 & 0.7 & 9.9142 & 3.8242 & 2.4189 & 1.5835 & 1.2305 & 1.1763 & 1.1376 & 1.0810 & 1.0153 & 0.9836 & 0.9625 \\
38.28 & 0.9 & 3.4557 & 2.0352 & 1.5906 & 1.2843 & 1.1380 & 1.1143 & 1.0941 & 1.0657 & 1.0322 & 1.0162 & 1.0067 \\
709 & 1 & 1.0122 & 1.0054 & 1.0036 & 1.0006 & 1.0005 & 1.0007 & 1.0003 & 1.0004 & 1.0001 & 1.0002 & 1.0001 \\

\bottomrule
\end{tabular}}}
\end{table}

\begin{table}[H]
\centering
\caption{MonteCarlo simulation: Super-additivity Ratio, Clayton copula}
\makebox[\textwidth]{
\scalebox{0.75}{
\begin{tabular}{cc *{11}{C}}

\toprule
q=5\% & $\alpha$ & \multicolumn{1}{c}{\textbf{0.2}} & \multicolumn{1}{c}{\textbf{0.3}} & \multicolumn{1}{c}{\textbf{0.4}} & \multicolumn{1}{c}{\textbf{0.6}} & \multicolumn{1}{c}{\textbf{0.9}} & \multicolumn{1}{c}{\textbf{1}} & \multicolumn{1}{c}{\textbf{1.1}} & \multicolumn{1}{c}{\textbf{1.3}} & \multicolumn{1}{c}{\textbf{1.6}} & \multicolumn{1}{c}{\textbf{1.8}} & \multicolumn{1}{c}{\textbf{2}} \\
\midrule

\multicolumn{1}{c}{Param. $\theta$ } & \multicolumn{1}{c}{ Kendall-$\tau$} & \multicolumn{2}{r}{\textbf{Clayton copula} } &  &   &  &   &   &   &   &   &   \\

\midrule
-1 & -1 & 5.8e-06 & 4.2e-06 & 3.3e-06 & 2.4e-06 & 1.8e-06 & 1.7e-06 & 1.6e-06 & 1.4e-06 & 1.3e-06 & 1.2e-06 & 1.1e-06 \\
-0.95 & -0.9 & 0.9751 & 0.5325 & 0.3778 & 0.2560 & 0.1832 & 0.1681 & 0.1562 & 0.1387 & 0.1223 & 0.1149 & 0.1083 \\
-0.82 & -0.7 & 3.0610 & 1.4229 & 0.9407 & 0.5954 & 0.4177 & 0.3842 & 0.3590 & 0.3208 & 0.2916 & 0.2801 & 0.2703 \\
-0.57 & -0.4 & 6.7280 & 2.6453 & 1.6368 & 0.9999 & 0.6877 & 0.6338 & 0.5909 & 0.5353 & 0.4923 & 0.4763 & 0.4625 \\
-0.46 & -0.3 & 8.5693 & 3.2123 & 1.9400 & 1.1556 & 0.7896 & 0.7265 & 0.6812 & 0.6152 & 0.5638 & 0.5416 & 0.5240 \\
-0.18 & -0.1 & 13.8704 & 4.5987 & 2.6015 & 1.4699 & 0.9976 & 0.9247 & 0.8652 & 0.7817 & 0.7071 & 0.6741 & 0.6479 \\
-0.02 & -0.01 & 15.7198 & 4.9752 & 2.8251 & 1.5697 & 1.0741 & 0.9948 & 0.9318 & 0.8455 & 0.7651 & 0.7291 & 0.7016 \\
0.00 & 0 & 16.0693 & 4.9913 & 2.8240 & 1.5876 & 1.0786 & 0.9981 & 0.9422 & 0.8518 & 0.7711 & 0.7348 & 0.7072 \\
0.02 & 0.01 & 15.6811 & 4.9562 & 2.7989 & 1.5823 & 1.0837 & 1.0037 & 0.9454 & 0.8576 & 0.7779 & 0.7424 & 0.7150 \\
0.22 & 0.1 & 12.3384 & 4.2727 & 2.5377 & 1.5152 & 1.0929 & 1.0222 & 0.9694 & 0.9004 & 0.8354 & 0.8054 & 0.7848 \\
0.50 & 0.2 & 7.9749 & 3.2764 & 2.1102 & 1.3780 & 1.0620 & 1.0143 & 0.9801 & 0.9312 & 0.8860 & 0.8654 & 0.8544 \\
0.86 & 0.3 & 4.9692 & 2.4106 & 1.7135 & 1.2439 & 1.0379 & 1.0030 & 0.9806 & 0.9532 & 0.9263 & 0.9154 & 0.9102 \\
1.33  & 0.4 & 3.2074 & 1.8440 & 1.4295 & 1.1446 & 1.0178 & 0.9989 & 0.9865 & 0.9716 & 0.9582 & 0.9512 & 0.9490 \\
4.67 & 0.7 & 1.3079 & 1.1236 & 1.0621 & 1.0201 & 1.0018 & 1.0000 & 0.9984 & 0.9966 & 0.9952 & 0.9945 & 0.9941 \\
18 & 0.9 & 1.0255 & 1.0099 & 1.0057 & 1.0016 & 0.9998 & 1.0001 & 0.9999 & 0.9997 & 0.9998 & 0.9995 & 0.9997 \\
2e06 & 1 & 1 & 1 & 1 & 1 & 1.0000 & 1 & 1 & 1 & 1.0000 & 1 & 1 \\
\bottomrule
\end{tabular}}}
\end{table}

\vspace*{-0cm}
\begin{table}[H]
\centering
\caption{MonteCarlo simulation: Super-additivity Ratio, Gumbel copula}
\makebox[\textwidth]{
\scalebox{0.75}{
\begin{tabular}{cc *{11}{C}}

\toprule
q=5\% & $\alpha$ & \multicolumn{1}{c}{\textbf{0.2}} & \multicolumn{1}{c}{\textbf{0.3}} & \multicolumn{1}{c}{\textbf{0.4}} & \multicolumn{1}{c}{\textbf{0.6}} & \multicolumn{1}{c}{\textbf{0.9}} & \multicolumn{1}{c}{\textbf{1}} & \multicolumn{1}{c}{\textbf{1.1}} & \multicolumn{1}{c}{\textbf{1.3}} & \multicolumn{1}{c}{\textbf{1.6}} & \multicolumn{1}{c}{\textbf{1.8}} & \multicolumn{1}{c}{\textbf{2}} \\
\midrule

\multicolumn{1}{c}{Param. $\theta$ } & \multicolumn{1}{c}{ Kendall-$\tau$} & \multicolumn{2}{r}{\textbf{Clayton copula} } &  &   &  &   &   &   &   &   &   \\

\midrule
1.00 & 0 & 16.0952 & 5.0635 & 2.8016 & 1.5854 & 1.0807 & 0.9984 & 0.9403 & 0.8519 & 0.7711 & 0.7349 & 0.7077 \\
1.01 & 0.01 & 16.0678 & 5.0694 & 2.8456 & 1.5901 & 1.0820 & 1.0034 & 0.9438 & 0.8545 & 0.7738 & 0.7386 & 0.7112 \\
1.11 & 0.1 & 15.9600 & 5.0510 & 2.8431 & 1.6124 & 1.1137 & 1.0306 & 0.9742 & 0.8887 & 0.8084 & 0.7738 & 0.7459 \\
1.25 & 0.2 & 15.2526 & 4.9554 & 2.8090 & 1.6174 & 1.1346 & 1.0620 & 1.0070 & 0.9236 & 0.8465 & 0.8094 & 0.7845 \\
1.43 & 0.3 & 14.0699 & 4.6658 & 2.7374 & 1.6078 & 1.1502 & 1.0810 & 1.0275 & 0.9535 & 0.8804 & 0.8460 & 0.8232 \\
1.67 & 0.4 & 12.4740 & 4.2964 & 2.5933 & 1.5756 & 1.1609 & 1.0940 & 1.0461 & 0.9807 & 0.9125 & 0.8800 & 0.8596 \\
2 & 0.5 & 10.1809 & 3.8287 & 2.3758 & 1.5114 & 1.1523 & 1.0994 & 1.0567 & 0.9995 & 0.9416 & 0.9122 & 0.8946 \\
3.33 & 0.7 & 5.1980 & 2.5551 & 1.8028 & 1.3201 & 1.1080 & 1.0779 & 1.0533 & 1.0180 & 0.9833 & 0.9662 & 0.9556 \\
5 & 0.8 & 3.1194 & 1.8558 & 1.4688 & 1.1913 & 1.0671 & 1.0495 & 1.0356 & 1.0163 & 0.9950 & 0.9842 & 0.9787 \\
10 & 0.9 & 1.6575 & 1.2853 & 1.1554 & 1.0637 & 1.0233 & 1.0195 & 1.0127 & 1.0060 & 0.9996 & 0.9964 & 0.9941 \\
100 & 0.99 & 1.0086 & 1.0043 & 1.0017 & 1.0009 & 1.0004 & 1.0002 & 1.0002 & 1.0003 & 1.0001 & 1.0000 & 0.9999 \\
1e06 & 1 & 1.0000 & 1 & 1 & 1.0000 & 1.0000 & 1 & 1.0000 & 1.0000 & 1.0000 & 1 & 1 \\
\bottomrule
\end{tabular}}}
\end{table}

\subsection{Diversification failure and tail dependence}

\begin{figure}[h]
 \begin{center}
\includegraphics[width=.45 \textwidth]{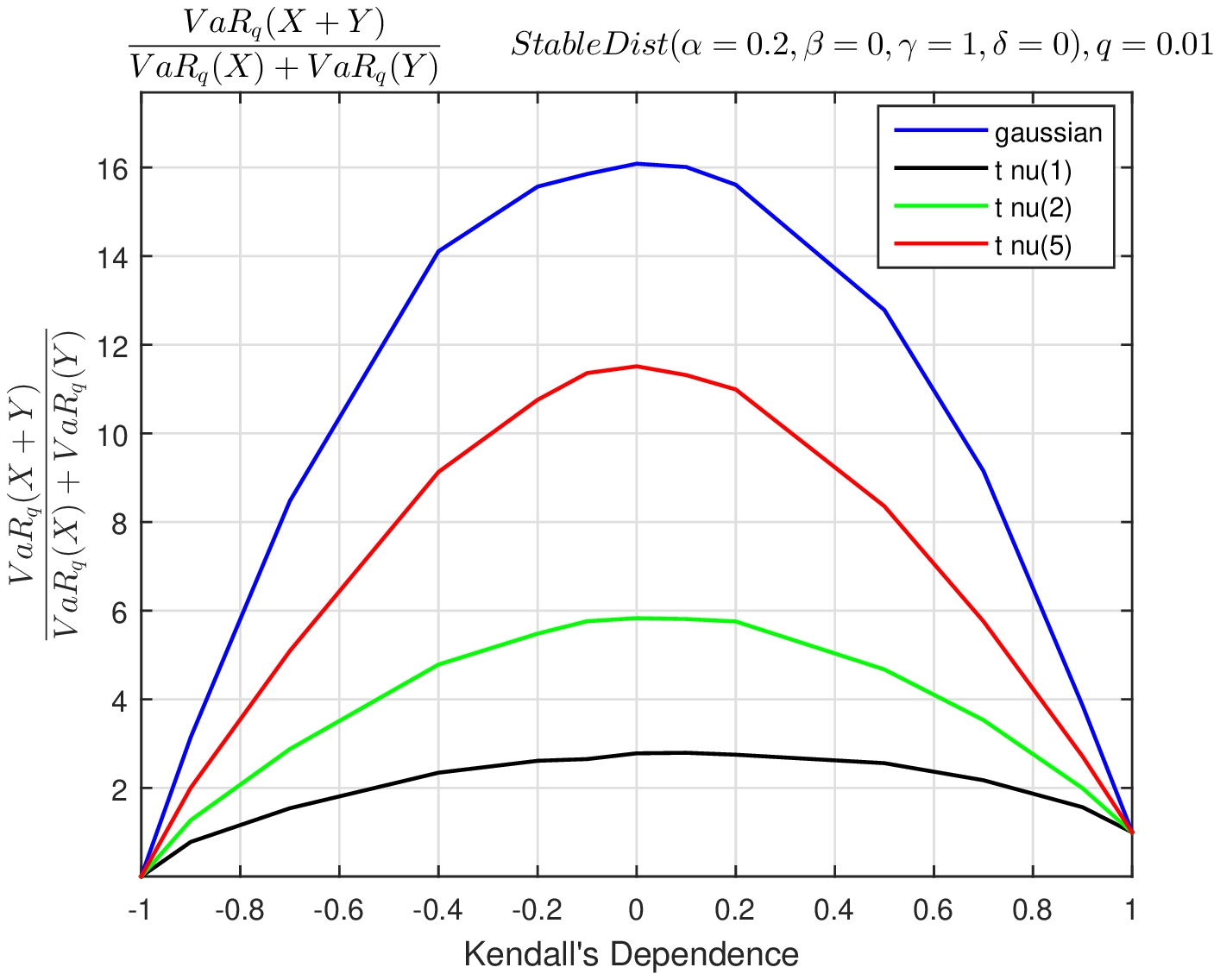} 
\includegraphics[width=.43 \textwidth]{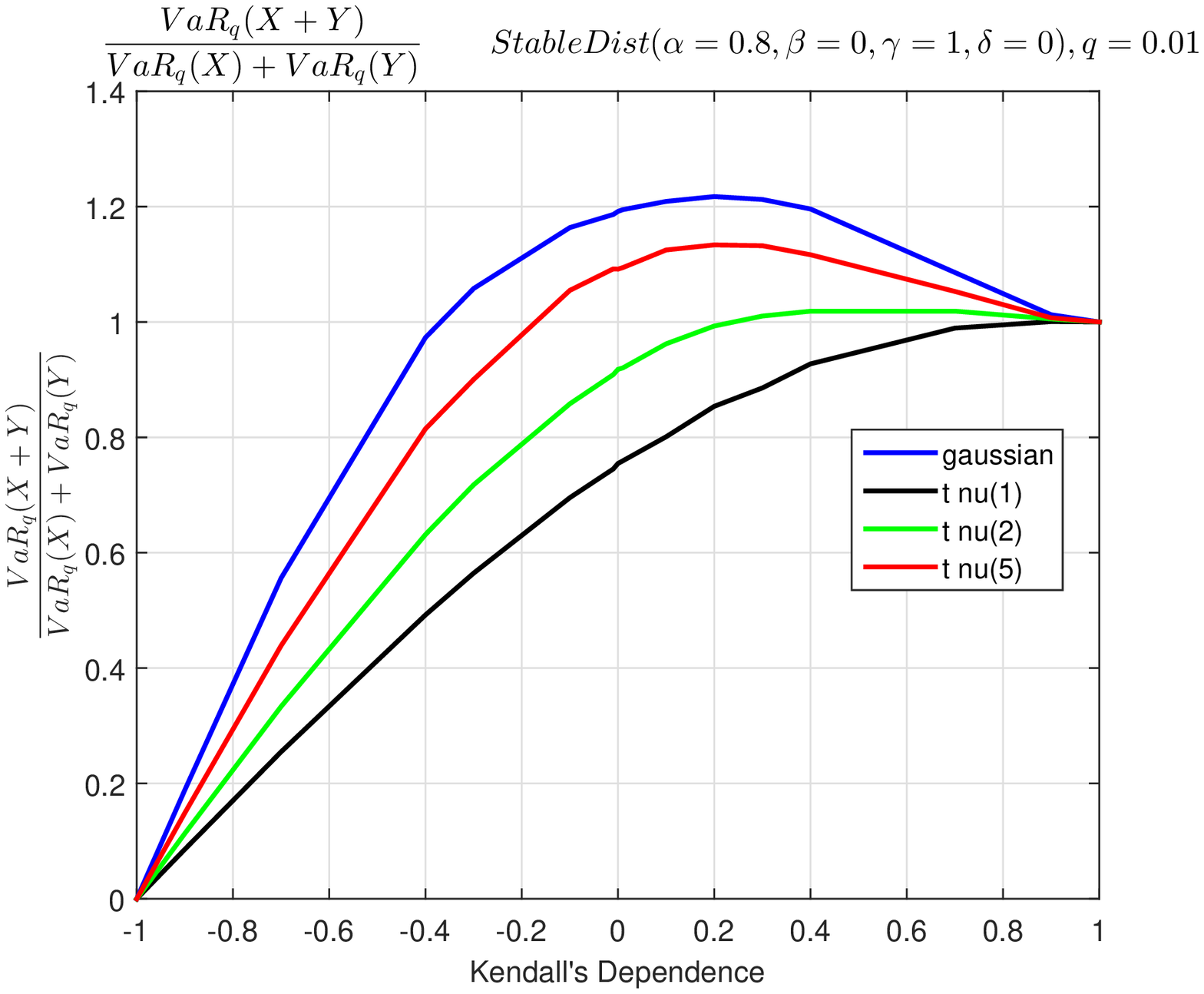}
\end{center}
\caption{Risk structure on different Stable marginals $(\alpha=0.2 , \alpha=0.8)$, elliptical copulas}
\label{fig:1}
\end{figure}

In this section we explore the impact of tail dependence and limits to diversification. The intuitive idea seems to be that higher tail dependence should worsen the scope for diversification. Surprisingly, our results point out exactly the opposite. In Figure \ref{fig:1} we report the graph of the relationship between the dependence and the super-additivity indexes for two different levels of $\alpha$, both in the region in which the first moment does not exist. The relationship is drawn for several copulas of the elliptical class with different tail-dependence indexes, that is the gaussian copula, for which the tail dependence is zero, and Student-t, for which the tail dependence increases for lower degrees of freedom. The highest tail dependence is then represented by the Student-t copula with $1$ degree of freedom.
We notice that for the case $\alpha =0.2$ the relationship is non monotone for all copulas, and the maximum of super-additivity is reached around the independence case. The interesting finding, that is new to the best of our knowledge, is that super-additivity is uniformly lower for copulas with higher tail dependence. The evidence is confirmed in the case $\alpha=0.8$, in which for the copula with highest tail dependence VaR is even sub-additive across the whole range of dependence.

So, there seems to be a puzzle in the result that higher tail dependence appears to ease the limits to diversification, rather than making them more binding. A very high degree of tail dependence may even destroy the non-monotonic nature of relationship between dependence and super-additivity, and so create a space for diversification.

\begin{figure}[h]
 \begin{center}
\includegraphics[width=.47 \textwidth]{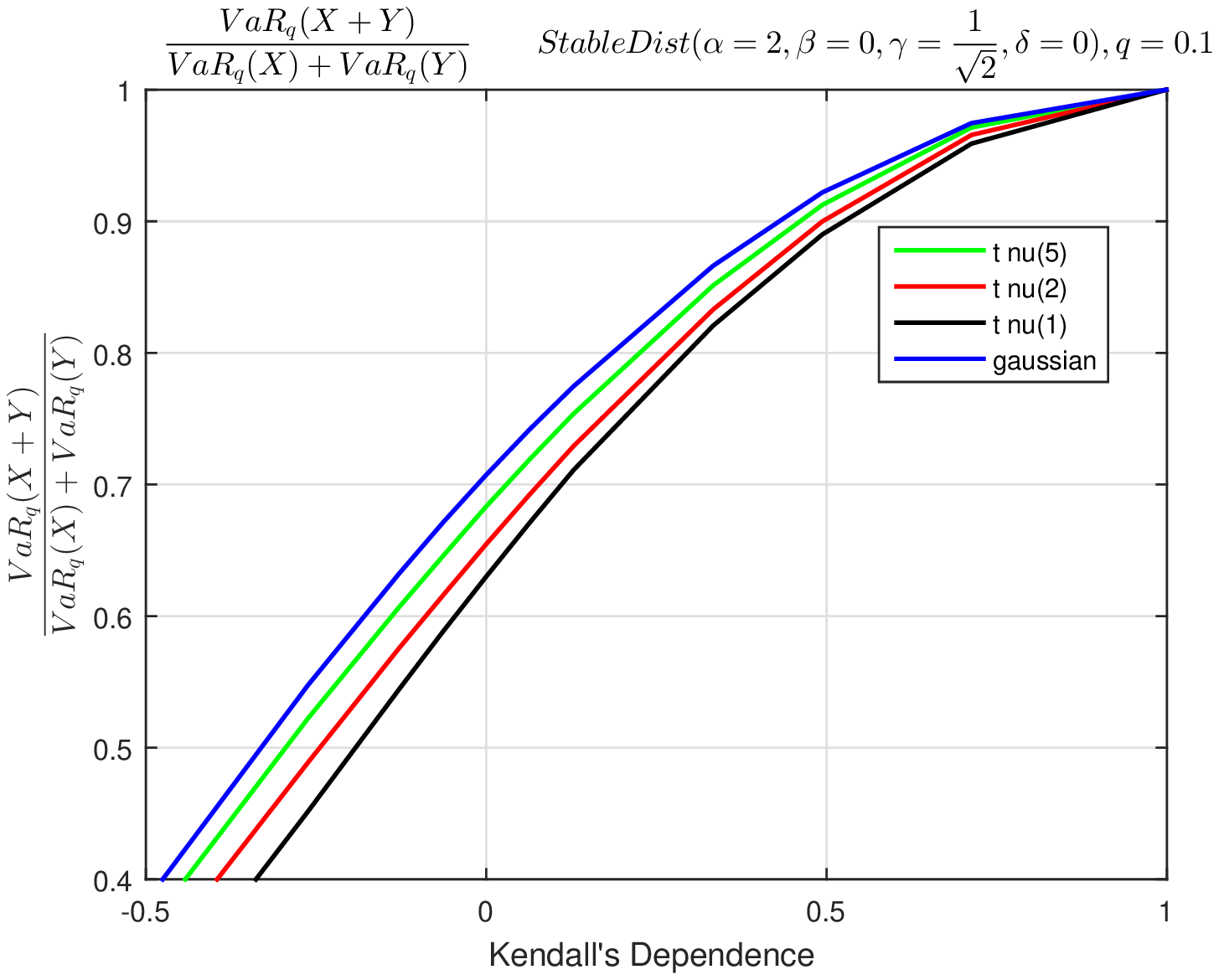}
\includegraphics[width=.45 \textwidth]{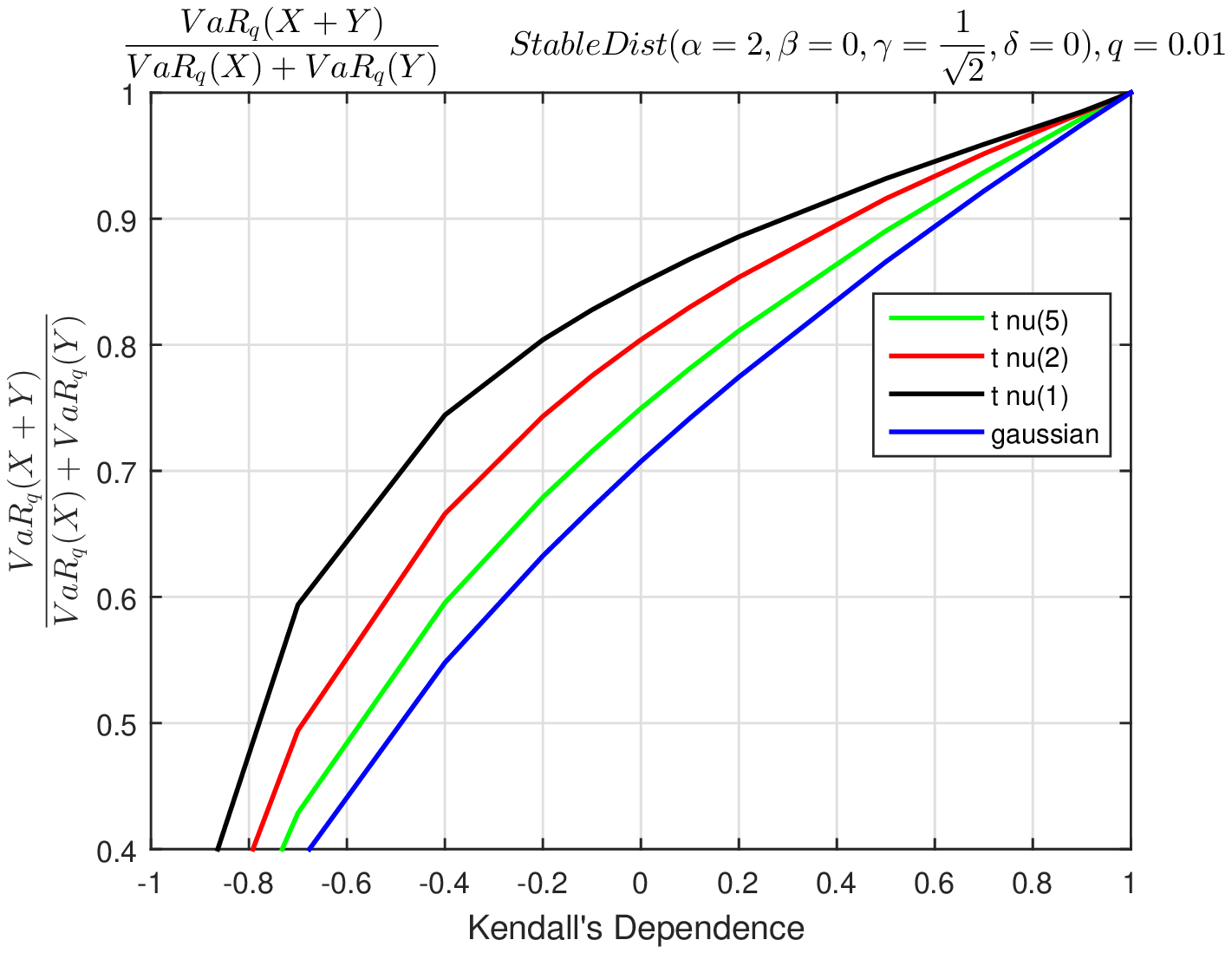}
\end{center}
\caption{ Risk structure on different quantile, Gaussian marginals, elliptical copulas}
\label{fig:3}
\end{figure}

The first conjecture that comes to mind to explain this result is that there could be a point in the tail where the expected ranking is established. The conjecture is suggested by what happens for gaussian risks, as documented in Figure \ref{fig:3}. In this case, the relationship between dependence and diversification shows that same ranking as that in the previous cases if we consider the $10\%$ percentile, while the order is completely reversed at the $1\%$ percentile.

\vspace*{-0cm}

\vspace*{0cm}

\begin{figure}[h]
 \begin{center}
\includegraphics[width=.46\textwidth]{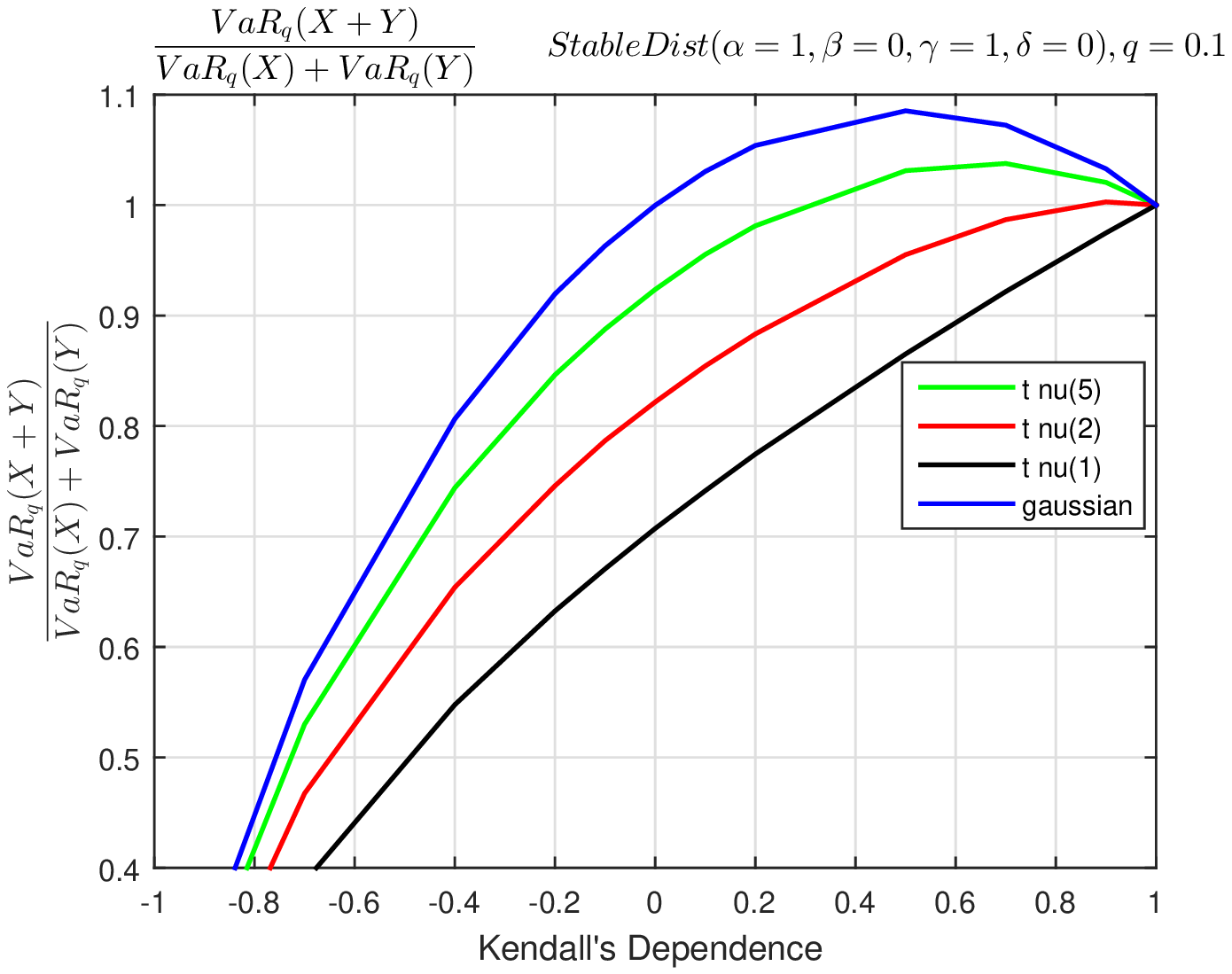}
\includegraphics[width=.46\textwidth]{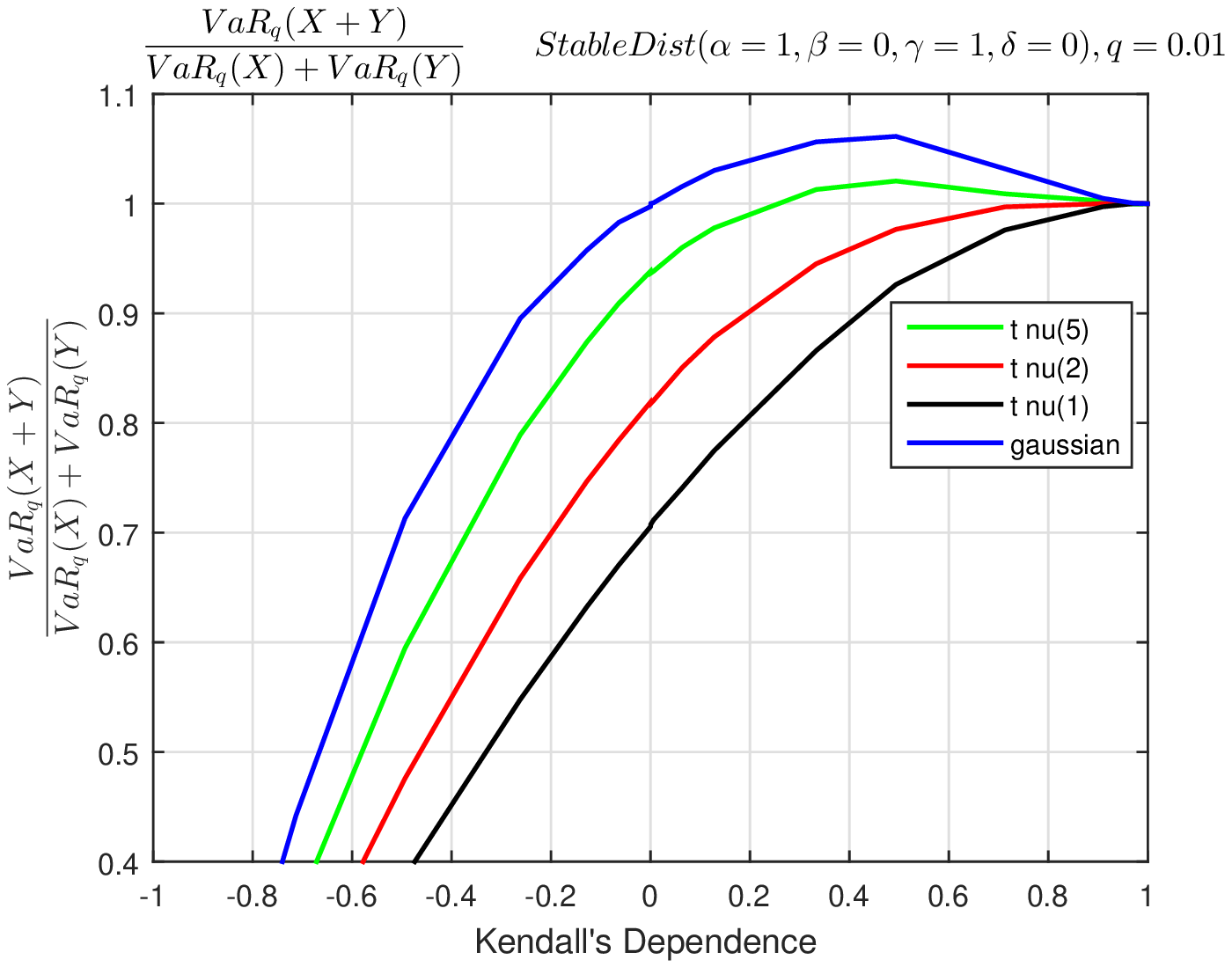}
\includegraphics[width=.48\textwidth]{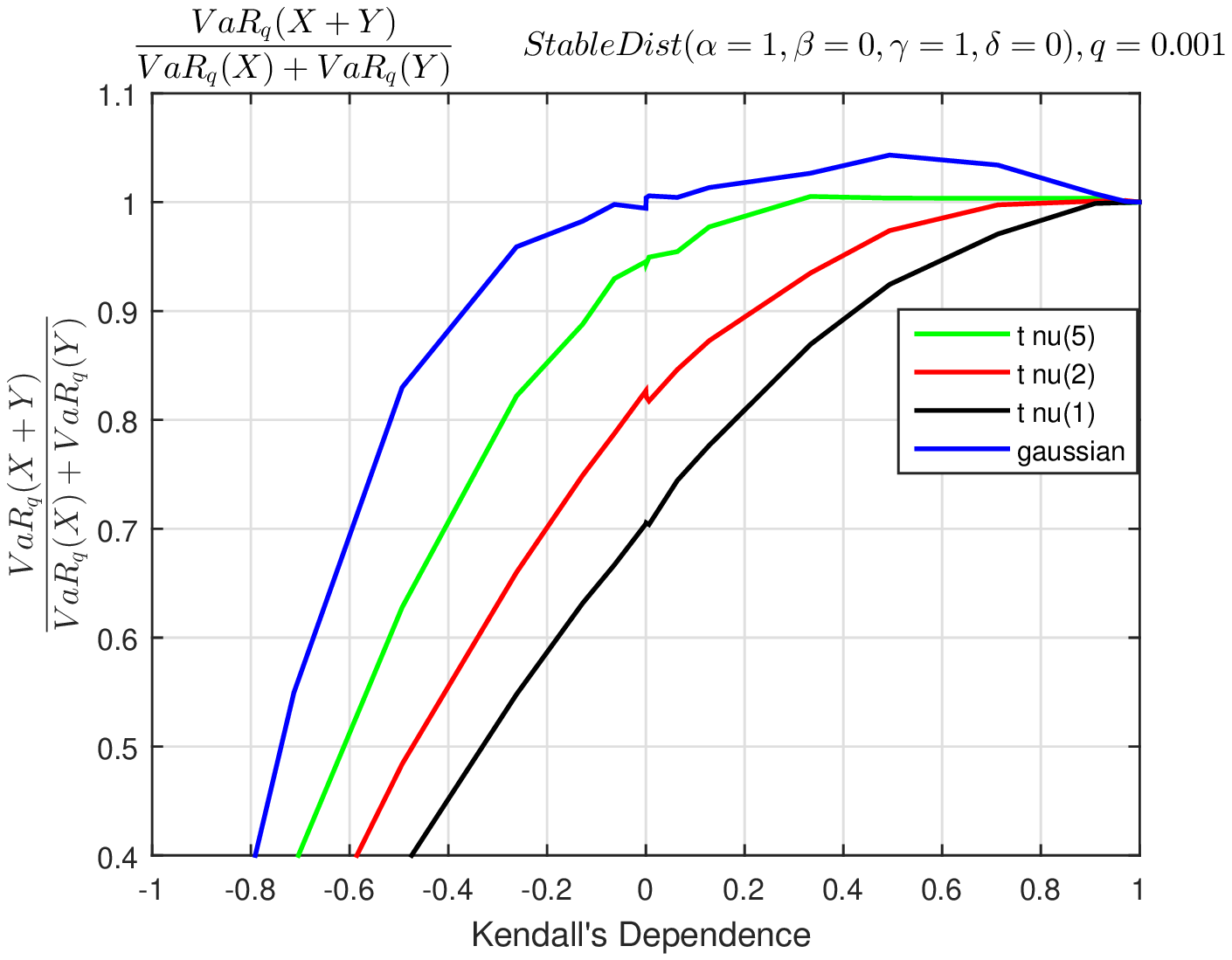}
\includegraphics[width=.47\textwidth, height=.36\textwidth]{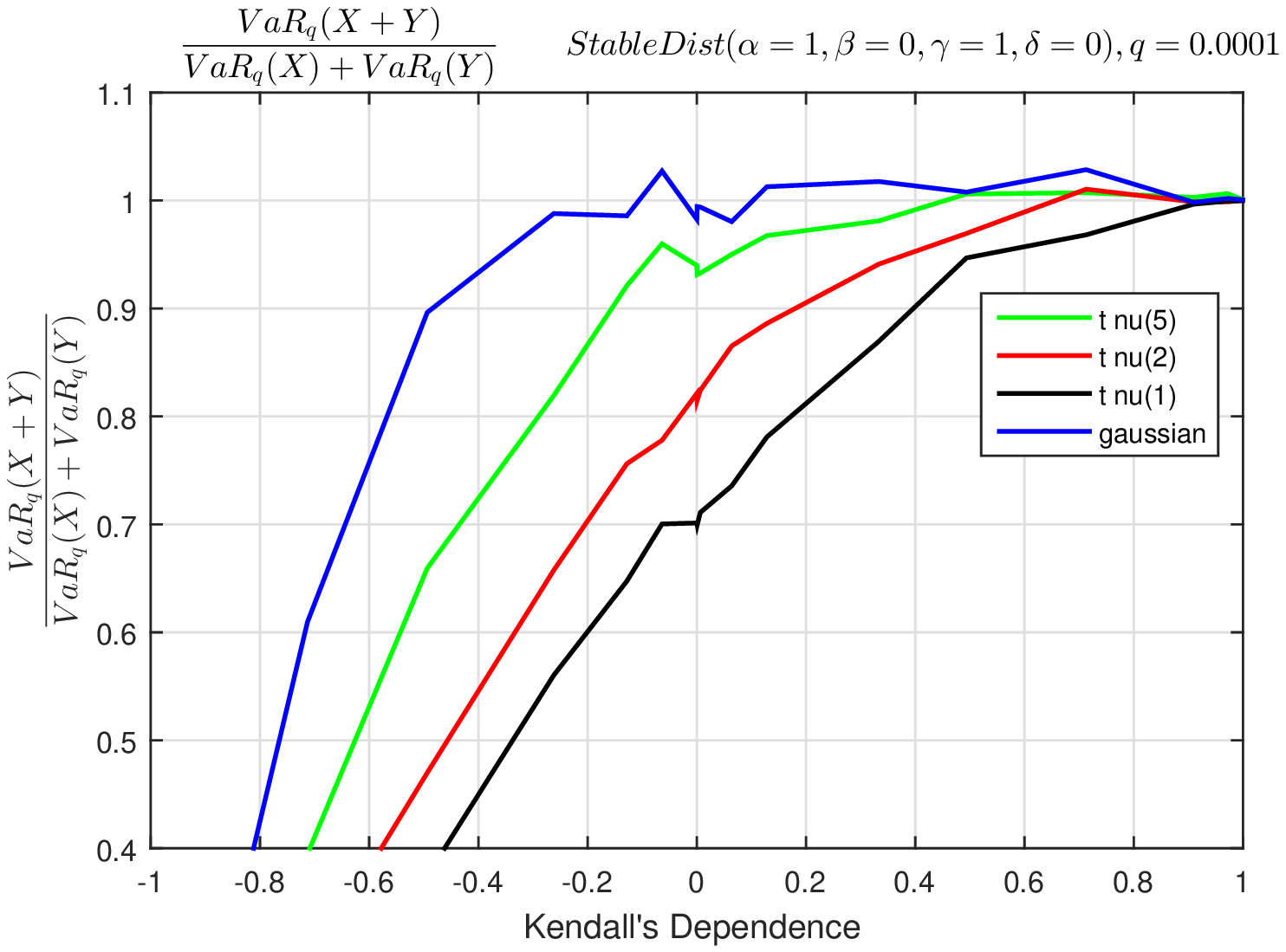}
\end{center}
\caption{Risk structure on different quantile, Cauchy marginals, elliptical copulas}
\label{fig:2}
\end{figure}

 Our conjecture is rejected if we perform the same analysis for Cauchy marginals ($\alpha=1$).
 We explore the relationships for percentiles far in the tail. (Figure \ref{fig:2}) shows that not only our conjecture is not confirmed, but a new facet of this tail dependence puzzle also emerges. Not only the schedule of the relationship between dependence and diversification is uniformly lower for copulas with increasing tail dependence. It also emerges that beyond some point in the tail the relationship is monotone. In other words, at some point in the tail a high degree of tail dependence creates diversification benefits. Even though the figure shows that the Monte Carlo error blurs the schedule of the relationship, the general path emerges quite clear. Deep in the tails, all the relationships become weakly monotonic, with an upper ceiling given by the comonotonic bound, $SR=1$, beyond a critical dependence level. Moreover, this additivity ceiling is reached for lower levels of dependence, the lower the tail dependence value of the copula function.

\vspace*{1.5cm}

\section{Conclusions}

    Heavy tailed $\alpha$-stable risks have been largely used in the representation of actuarial risks, particularly in the field of catastrophe insurance, and operational risk, which is the kind of financial risk which is closest to the kinds of risk that are typical of insurance. In this paper, we provide simulations to study the aggregate behavior of losses for $\alpha$-stable risks, with particular reference to a \emph{super-additivity} index.

    We contribute to the literature on this kind of risk in two directions. For what concerns the insurance market, we support evidence concerning the impossibility of re-insurance policies. We document that any reinsurance policy would call for more capital than that needed to shield the losses on individual risks. This effect is particularly evident for risks with low $\alpha$, that is very heavy tails, and positive dependence. Moreover, the super-additivity curse mainly hits cases in which the $\alpha$ parameter is lower than $1$ and the mean is not defined. This implies that in this case coherent risk measures like the \emph{Expected Shortfall} are not available and the Value-at-Risk is the obliged choice.

    As for the analysis of diversification in the tail of the distribution, our simulations provide evidence that in our view represents a new puzzle. If we rank the dependence structures according to their tail dependence indexes, we find that dependence structures with higher tail dependence provide more space for diversification. In particular we find that: i) the super-additivity ratio for every level of dependence is lower for dependence structures with higher tail dependence; ii) for extreme points in the tail the aggregate VaR of risks becomes additive beyond a critical dependence level, and this critical level is lower for copula functions with lower tail dependence.

    Concerning the latter point, our simulation approach opens a new research path that applies simulation analysis to explore the very extreme tail behaviour of aggregated risks, parallel to a stream of literature that instead has studied the asymptotic tail behaviour of aggregated VaR from an analytical point of view (see for example \citet{embrechts2009additivity}). A more extensive empirical investigation of the actual convergence of the aggregated VaR to this limit is left as a topic for future research.

    For what instead concerns the policy implications of our findings, the open question to be addressed is whether there is any alternative to the reinsurance choice that could allow to transfer heavy tailed $\alpha$-stable risks and to create room for diversification. The natural alternative is of course securitization, mixing small percentages of $\alpha$-stable risks in very well diversified financial portfolios. This alternative would also deserve to be analyzed in depth with convolution analysis and simulations, and if it were to fail, that would leave the general public, and so the taxpayer, as the only possible reinsurance counter party for this kind of risk.

   \bibliographystyle{apalike}
   \bibliography{biblio}   


\end{document}